\documentclass[letterpaper, 10 pt, conference]{ieeeconf}  % Comment this line out if you need a4paper

\IEEEoverridecommandlockouts                              % This command is only needed if 
\usepackage{amsmath,amssymb,amsfonts}
\usepackage{algorithmic}
\usepackage{graphicx}
\usepackage{textcomp}
\usepackage{subfigure}
\usepackage{tabularx}
\usepackage{array}
\usepackage{stfloats}
\usepackage{multirow}
\usepackage{xcolor}

\makeatletter
\let\NAT@parse\undefined
\makeatother

\usepackage[hidelinks,breaklinks=true]{hyperref}

\title{\LARGE \bf
Subject-Specific Analysis of Self-Initiated Attention Shifts from EEG with Controlled Internal and External Attention Conditions
}

\author{
Yuwen Zeng$^{1}$, \textit{Member, IEEE},
Dengzhe Hou$^{2,\dagger}$,
Zhang Zhang$^{3}$, \textit{Member, IEEE},\\
Sai Sun$^{4}$,
Yongsong Huang$^{1}$, \textit{Member, IEEE},
Chia-huei Tseng$^{4}$,
and Satoshi Shioiri$^{1,\dagger}$%
\thanks{*This work was supported by JSPS KAKENHI (Grant No. JP25KJ0641 and JP24H00700).}%
\thanks{$^{\dagger}$Corresponding authors: Dengzhe Hou and Satoshi Shioiri.}%
\thanks{$^{1}$Yuwen Zeng, Yongsong Huang, and Satoshi Shioiri are with the Advanced Institute of Convergence Knowledge Informatics, Tohoku University, Japan 
{\tt\small (yuwen@tohoku.ac.jp; huang.yongsong.c5@tohoku.ac.jp; satoshi.shioiri.b5@tohoku.ac.jp)}}%
\thanks{$^{2}$Dengzhe Hou is with the Graduate School of Information Sciences, Tohoku University, Japan 
{\tt\small dengzhe.hou.a5@tohoku.ac.jp}}%
\thanks{$^{3}$Zhang Zhang is with the Center for Data-Driven Science and Artificial Intelligence, Tohoku University, Japan 
{\tt\small zhangzhang@tohoku.ac.jp}}%
\thanks{$^{4}$Sai Sun and Chia-huei Tseng are with the Research Institute of Electrical Communication, Tohoku University, Japan 
{\tt\small (sunsai1215@gmail.com; chia-huei.tseng.a8@tohoku.ac.jp)}}%
}

\begin{document}

\maketitle
\thispagestyle{empty}
\pagestyle{empty}

%%%%%%%%%%%%%%%%%%%%%%%%%%%%%%%%%%%%%%%%%%%%%%%%%%%%%%%%%%%%%%%%%%%%%%%%%%%%%%%%
\begin{abstract}

Self-initiated attention shifts play a critical role in voluntary behavior but are difficult to study due to the absence of explicit temporal markers. While previous studies have examined their neural correlates, it remains unclear how multi-dimensional electroencephalography (EEG) features contribute to their characterization within an interpretable computational framework. In this study, we build on an experimental paradigm developed in our previous work, which enables controlled comparison between task-constrained self-initiated shifts and externally instructed shifts under identical visual stimulation. Within this setting, we investigate whether preparatory EEG activity can distinguish these two types of attention shifts. We adopt a machine learning–based approach and conduct two complementary analyses: (1) a performance-oriented assessment of frequency-specific topographic patterns, and (2) a model-based feature attribution analysis using SHapley Additive exPlanations (SHAP). These analyses provide a structured view of how spectral features across regions of interest contribute to model behavior. Our results demonstrate reliable within-subject classification performance, indicating that preparatory EEG activity contains subject-specific discriminative information within this paradigm. The analysis shows that higher-frequency bands and frontal regions contribute strongly to model decisions, although such contributions should be interpreted cautiously due to the potential influence of non-neural artifacts in high-frequency EEG signals. Overall, this work highlights the value of interpretable machine learning for analyzing subject-specific EEG signal patterns in a controlled experimental setting, with potential applications in personalized and asynchronous brain–machine interface systems.

\end{abstract}

%%%%%%%%%%%%%%%%%%%%%%%%%%%%%%%%%%%%%%%%%%%%%%%%%%%%%%%%%%%%%%%%%%%%%%%%%%%%%%%%
\section{Introduction}

Human attention can be guided by external stimuli or initiated internally based on goals and intentions \cite{PetersenPosner}. Externally cued attention has been widely studied using stimulus-locked paradigms, where neural responses are aligned to well-defined events. In contrast, self-initiated shifts do not have explicit timing markers, which makes them difficult to analyze using the same approach \cite{NadraReview}.

Methods such as Steady-State Visually Evoked Potentials (SSVEPs) provide a reliable way to study attentional allocation by tracking neural synchronization to external rhythmic stimulation \cite{Kashiwase2012, Shioiri2016}. However, these paradigms mainly reflect responses driven by external input. Internally generated shifts are less accessible in such settings, especially when timing is not externally constrained. This has led to increasing interest in neural activity preceding attention shifts. Because self-initiated shifts occur without explicit triggers, standard averaging methods are often insufficient \cite{NadraWilled, WuSI1, WuSI2}. Single-trial analysis of preparatory activity is therefore becoming important, both for understanding voluntary attention and for asynchronous brain–machine interface (BMI) systems.

Previous studies have linked voluntary attention to oscillatory activity in EEG. Frontal theta is often associated with cognitive control, while posterior alpha is related to spatial orienting and target detection \cite{ThutAlpha, WordenAlpha, ClementsThetaAlpha, HouTheta}. These findings indicate that different frequency bands are involved in different aspects of attention. At the same time, most analyses rely on group-level averaging and predefined hypotheses, which may not capture variability across individuals. There is also evidence that spontaneous neural fluctuations before a decision can carry predictive information about internally generated attention \cite{Bengson2014}. This preparatory activity has been described as a stochastic accumulation process \cite{Schurger2012}, suggesting that the period before a shift contains subject-specific information that is not fully reflected in conventional analyses.

Recent work, including our previous study \cite{HouTheta}, has addressed some of these limitations by introducing an experimental design that allows flexible shift timing and multiple choice options under controlled visual input. This design enables a structured and unique comparison between internally and externally driven attention shifts under matched conditions, while maintaining identical visual stimulation.

Machine learning methods are increasingly used to analyze EEG data in both research and BMI applications \cite{SaeidiReview,SunAttention,LotteBCIClassif}. These methods can handle high-dimensional features and capture complex patterns, but their behavior is often difficult to interpret. This makes it harder to relate model outputs to meaningful signal characteristics \cite{FellousXAI,CaoBCIXAI}. Techniques such as SHAP provide a way to summarize how features contribute to model output \cite{LundbergSHAP}. In EEG analysis, this can be used to organize contributions across frequency bands, spatial regions, and feature types.

In this work, we build on this experimental paradigm to analyze preparatory EEG activity associated with task-constrained self-initiated and externally instructed attention shifts. The dataset provides a structured setting that extends beyond standard cue-based designs while maintaining controlled experimental conditions. Within this setting, we focus on subject-specific analysis and examine whether consistent patterns can be identified within individuals, rather than emphasizing cross-subject generalization. To characterize these patterns, we combine a performance-oriented analysis of frequency-specific topographic distributions with a SHAP-based analysis of feature contributions. This approach allows us to examine how spectral features across different regions of interest (ROI) are utilized by the model, providing an interpretable description of subject-specific EEG signal patterns. The resulting framework may support the development of personalized and asynchronous BMI systems.

The main contributions of this work can be summarized as follows:
(1) We present a subject-specific analysis of preparatory EEG activity based on a recently introduced experimental paradigm from our prior work, which enables controlled comparison between internally and externally driven attention shifts under identical visual stimulation.
(2) Within this framework, we combine frequency-specific topographic analysis with SHAP-based feature attribution to provide an interpretable characterization of how spectral features and spatial regions are utilized in distinguishing attention states.

\section{Method}

\subsection{Dataset and Experimental Paradigm}
The analysis is based on an EEG dataset collected in our previous work using a rapid serial visual presentation (RSVP) paradigm \cite{HouTheta}. The experimental design, EEG acquisition, and preprocessing procedures follow those described in that study. Fifteen healthy adult participants took part in the experiment. Due to individual differences in the tendency to shift attention, the number of valid trials after eye-tracking and artifact rejection varied across participants, ranging from a few tens to more than one hundred trials. To account for this variability, stratified cross-validation was applied within each participant to preserve class distributions.

Participants viewed four RSVP streams presented simultaneously at peripheral locations. The analysis focuses on two task conditions: task-constrained self-initiated (TCSI) shifts and externally instructed (EI) shifts. While both conditions involve goal-directed behavior, they differ in whether the intention to shift is generated internally or externally. Visual stimulation was identical across conditions, and attention shifts were identified based on eye-tracking data. The free-viewing condition was excluded to maintain consistent task structure and reduce variability in cognitive state. Restricting the analysis to task-constrained shifts allows a controlled comparison between internally and externally initiated attention under matched conditions.

\subsection{EEG Segmentation and Band-Specific Representations}
EEG preprocessing was conducted according to the pipeline described in the original study \cite{HouTheta}. For each attention shift, EEG segments were extracted from $-2.0$ s to $-0.5$ s relative to the shift onset. This time window was selected to capture preparatory activity while reducing potential contamination from eye-movement artifacts.

Time-frequency representations were used to derive EEG features for five frequency bands: theta (4–7 Hz), alpha (8–12 Hz), low beta (13–20 Hz), high beta (20–30 Hz), and gamma (30–40 Hz). Two input configurations were considered. In the single-band setting, features were extracted independently from each frequency band. In the multi-band setting, features from all bands were concatenated to form a combined representation. This design allows assessment of whether discriminative information is concentrated within specific frequency bands or distributed across multiple bands.

\subsection{Feature Extraction}
Based on the band-specific representations, a feature set was constructed to capture different aspects of the EEG signal. The features were grouped into three categories. The first includes global features that describe overall signal statistics without reference to specific regions. The second consists of intra-ROI features that characterize spatial variability and temporal dynamics within individual regions of interest (ROIs). The third includes inter-ROI features that describe relationships between regions, including Pearson correlation-based connectivity and hemispheric asymmetry.

\begin{figure}[!tb]
\centerline{\includegraphics[width=4cm]{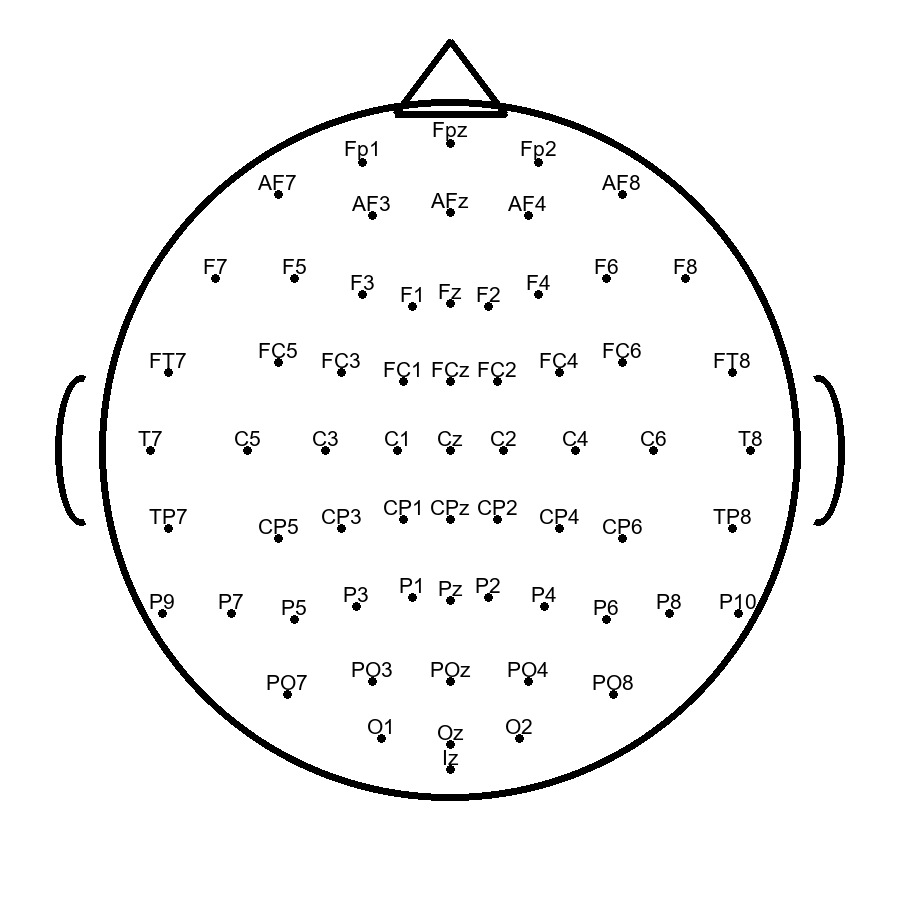}}
\caption{Electrode distribution of the BioSemi 64-channel montage. This figure shows a schematic 2D projection of the electrode layout.}
\label{fig_electrode}
\end{figure}

EEG was recorded using a 64-channel BioSemi system, and electrode labels follow the standard BioSemi 64 montage. The electrode layout is shown in Fig.~\ref{fig_electrode}, and ROI definitions are provided in Table~\ref{tab_electrode}.

\begin{table}[!tb]
\renewcommand{\arraystretch}{1.3}
\centering
\caption{ROI definitions of electrode groups}
\label{tab_electrode}
\setlength{\tabcolsep}{1.5pt}
\begin{tabularx}{\linewidth}{>{\raggedright\arraybackslash}p{1.12cm}|*{3}{>{\raggedright\arraybackslash}X}}
\hline
\textbf{ROI} & \textbf{Left} & \textbf{Right} & \textbf{Midline} \\
\hline
Prefrontal & Fp1, AF7, AF3, F1 & Fp2, AF8, AF4, F2 & -- \\
Frontal & F3, F5, F7, FT7 & F4, F6, F8, FT8 & Fpz, AFz, Fz, FCz \\
Fronto-central & FC5, FC3, FC1, C1 & FC6, FC4, FC2, C2 & -- \\
Centro-temporal & C3, C5, T7, TP7 & C4, C6, T8, TP8 & -- \\
Parietal1 & CP5, CP3, CP1, P1 & CP6, CP4, CP2, P2 & Cz, CPz, Pz, POz \\
Parietal2 & P3, P5, P7, P9 & P4, P6, P8, P10 & -- \\
Occipital & PO7, PO3, O1, Iz & PO8, PO4, O2, Oz & -- \\
\hline
\end{tabularx}
\end{table}

In the multi-band setting, concatenation of features from all frequency bands resulted in more than $5{,}000$ features per trial. To address the high dimensionality relative to the number of trials, a stratified feature selection procedure was applied within each training fold. Features were ranked using ANOVA F-scores within each band and feature type. To avoid information leakage, ANOVA-based feature selection was performed separately within each cross-validation split using only the training data. The selected feature indices were then applied to the held-out fold, which was not used for feature ranking, model fitting, or hyperparameter determination. A subset of top-ranked features was then selected from global ($15\%$), intra-ROI spatial-variability ($30\%$), intra-ROI temporal-dynamics ($30\%$), and inter-ROI connectivity ($25\%$) categories, resulting in a total of $500$ features. The same procedure was applied in the single-band setting with $100$ features retained. These ratios were used as a predefined design choice to ensure that all major feature categories were represented in the final feature set. They were not optimized as model hyperparameters, and the resulting SHAP contributions should therefore be interpreted as properties of this specific feature representation.

\subsection{Classification and Analysis}
A binary classification task was defined to distinguish TCSI shifts from EI shifts. A Random Forest classifier was applied due to its robustness in high-dimensional settings and its compatibility with feature importance analysis \cite{SaeidiReview,KimMuBeta,LuRFMotor}. The model used 200 trees, a maximum depth of 10, and a minimum of 10 samples per split, with balanced class weights. These hyperparameters were selected in preliminary testing and then kept fixed across all cross-validation folds and conditions.

To account for inter-individual variability, a within-subject evaluation scheme was adopted. Separate models were trained for each participant using three-fold stratified cross-validation. Classification performance was evaluated using accuracy and area under the ROC curve (AUC). In addition, cross-subject generalization was explored using a leave-one-subject-out setting.

Two additional analyses were conducted to characterize model behavior. First, time-averaged topographic patterns were computed by averaging band-specific EEG power over the pre-shift interval. Second, SHAP was used to quantify the contributions of different feature types and ROIs to model outputs \cite{LundbergSHAP}. This analysis provides a structured summary of feature contributions to model behavior rather than a direct interpretation of underlying neural mechanisms.

\section{Results}
\subsection{Classification Performance}
\begin{figure}[!tb]
\centerline{\includegraphics[width=8cm]{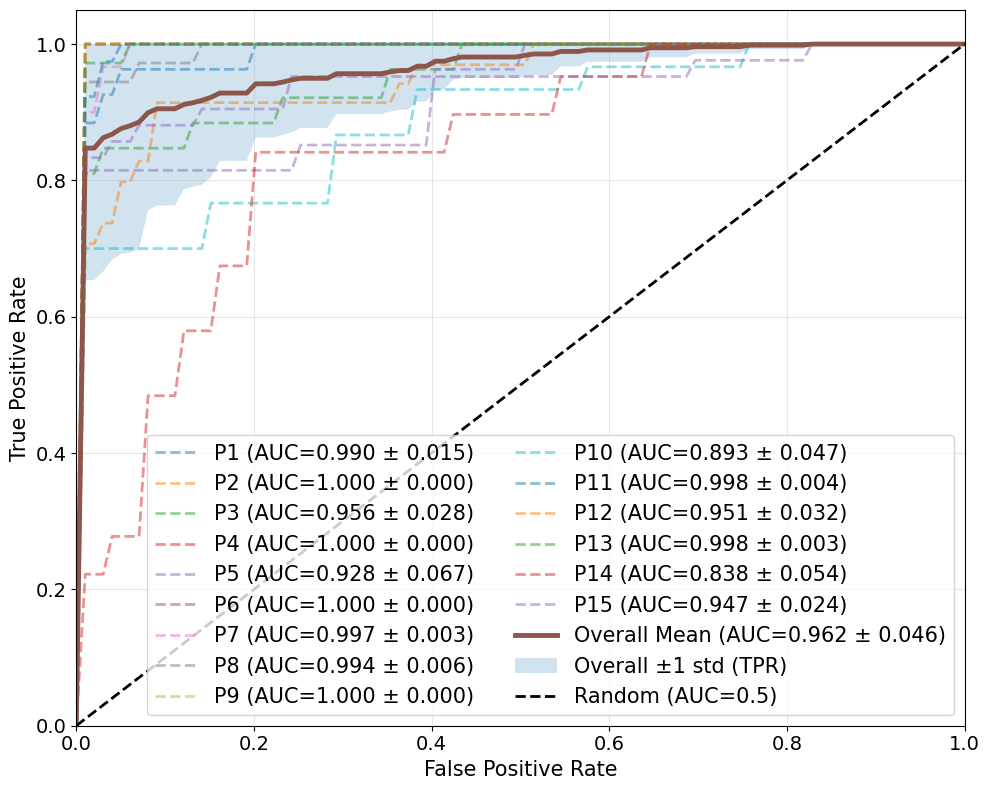}}
\caption{ROC curves and AUC values for within-subject classification between TCSI and EI shifts using the multi-band model, where the blue shaded area indicates $\pm 1$ standard deviation around the mean ROC curve.}
\label{fig_roc}
\end{figure}

Fig. \ref{fig_roc} illustrates the receiver operating characteristic (ROC) curves and AUC values for the within-subject classification of TCSI and EI shifts using the multi-band model. For brevity, the performance metrics for both multi-band and single-band configurations are summarized in Table \ref{tab2}. The multi-band and gamma-only models achieved comparable performance, with AUC values of $0.962 \pm 0.046$ and $0.964 \pm 0.038$, respectively. Thus, the role of the multi-band representation in this study is not to provide a clear performance advantage, but to provide a unified feature space for comparing band-level, feature-type-level, and ROI-level contributions.

These results suggest that higher-frequency components provide strong discriminative signals for the classification task. However, such contributions should be interpreted with caution, as high-frequency EEG activity may be influenced by residual muscular or oculomotor artifacts. Across all evaluated settings, average AUC values remained substantially above the chance level, indicating that preparatory EEG activity contains sufficient information to differentiate TCSI shifts from EI shifts at the within-subject level. In contrast, cross-subject decoding using a leave-one-subject-out scheme remained at chance level, highlighting substantial inter-individual variability. This observation further supports the importance of subject-specific modeling in this context. Consequently, the subsequent topographic and SHAP analyses focus on within-subject models.

\begin{table*}[!tb]
\renewcommand{\arraystretch}{1.3}
\centering
\caption{Within-subject classification accuracy and AUC for models trained with multi-band and single-band feature representations}
\label{tab2}
\begin{tabular}{l|llllll}
\hline
\multirow{2}{*}{Metric} & \multirow{2}{*}{Multi-band} & \multicolumn{5}{c}{Single-band}                                                         \\ \cline{3-7} 
                        &                            & Theta           & Alpha           & Low   Beta      & High   Beta     & Gamma           \\ \hline
Accuracy                & 0.936   ± 0.046            & 0.782   ± 0.122 & 0.793   ± 0.125 & 0.880   ± 0.085 & 0.915   ± 0.061 & 0.935   ± 0.054 \\
AUC                     & 0.962   ± 0.046            & 0.781   ± 0.161 & 0.798   ± 0.144 & 0.913   ± 0.080 & 0.948   ± 0.045 & 0.964   ± 0.038 \\ \hline
\end{tabular}
\end{table*}

\subsection{Band-Specific Topographic Importance Analysis}
To assess the predictive relevance of different frequency bands, we conducted a performance-oriented analysis based on spatial patterns. For each frequency band, EEG power was averaged over the pre-shift time window and projected onto the scalp to obtain time-averaged topographic maps. For each ROI, band-specific models were trained using features restricted to that ROI, and the resulting AUC was used as a regional importance score for visualization.

\begin{figure}[!b]
\centerline{\includegraphics[width=8cm]{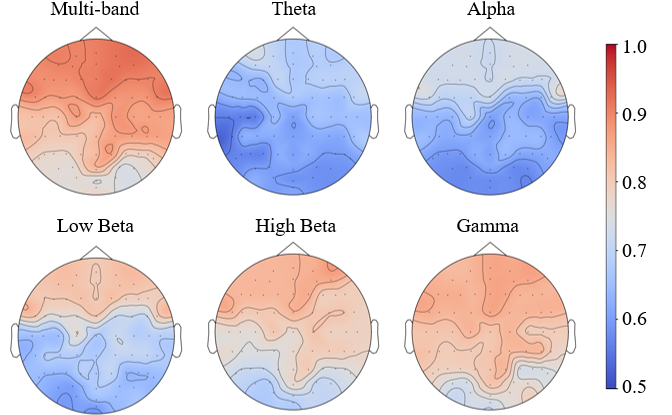}}
\caption{Topographic patterns for the multi-band setting and the five individual frequency bands, averaged over the pre-shift interval. Colors indicate the AUC-based importance of each ROI in the classification of TCSI and EI.}
\label{fig_bandtopo}
\end{figure}

Fig. \ref{fig_bandtopo} shows the resulting topographic patterns for the multi-band setting and for the five individual bands. Distinct spatial distributions are observed across frequency bands. Higher-frequency bands, particularly gamma and high beta, exhibit more structured and focal patterns, whereas lower-frequency bands show more diffuse distributions.

These differences in spatial structure are consistent with the corresponding classification performance. However, rather than indicating that specific frequency bands are exclusively responsible for the underlying cognitive process, these results suggest that different bands provide complementary information for the classification model under the current feature representation.

\subsection{SHAP-Based Model Explanation}
To investigate how the multi-band model integrates information across the frequency spectrum, we utilized SHAP to quantify feature-level contributions. 

\begin{figure}[!tb]
\centerline{\includegraphics[width=7.5cm]{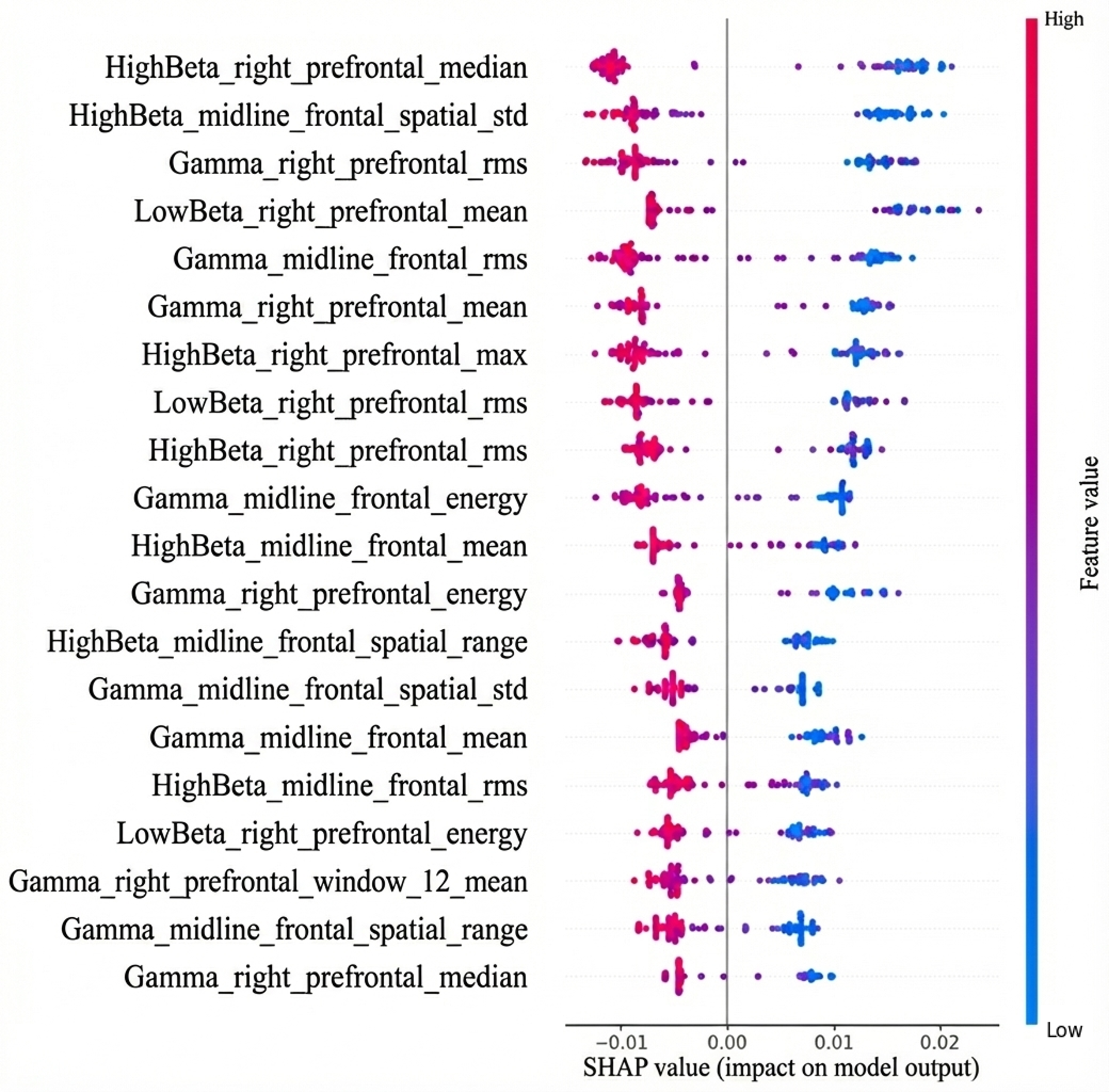}}
\caption{Example SHAP summary plot for a representative participant in the multi-band setting; each point indicates the contribution of a feature to the model output for individual trials, and negative SHAP values (left) favor TCSI while positive SHAP values (right) favor EI.}
\label{fig_shap}
\end{figure}

SHAP values were aggregated within each feature group to account for differences in feature dimensionality. Absolute SHAP values were normalized per participant so that contributions sum to one at each level (band, feature type, and ROI). Fig. \ref{fig_shap} presents an example SHAP summary plot for a representative participant.

At the band level (Table~\ref{tab:shap_band}), higher-frequency bands show larger normalized contributions, with gamma and high beta exhibiting the highest mean $|\mathrm{SHAP}|$ values. Low beta contributes at an intermediate level, while alpha and theta show smaller contributions. The relatively large standard deviation for gamma indicates notable variability across participants, suggesting that the dominance of high-frequency features is not uniform across individuals.

\begin{table}[!tb]
\renewcommand{\arraystretch}{1.3}
\centering
\caption{Band-level normalized SHAP contributions in the within-subject multi-band model. The standard deviation (SD) indicates inter-participant variability.}
\label{tab:shap_band}
\begin{tabular}{lcc}
\hline
Band & Mean norm. $|$SHAP$|$ & SD \\
\hline
Gamma ($20\%$)   & 0.367 & 0.129 \\
HighBeta ($20\%$)& 0.286 & 0.073 \\
LowBeta ($20\%$) & 0.165 & 0.069 \\
Alpha ($20\%$)   & 0.103 & 0.072 \\
Theta ($20\%$)   & 0.080 & 0.062 \\
\hline
\end{tabular}
\end{table}

Importantly, these SHAP values reflect the contribution of features within the constructed representation and should not be interpreted as direct evidence of the relative importance of underlying neural oscillations. In particular, the lower contribution of theta does not contradict prior findings in cognitive neuroscience, but rather indicates that, under the present feature design and classification setting, higher-frequency features provide stronger discriminative signals.

\begin{table}[!tb]
\renewcommand{\arraystretch}{1.3}
\centering
\caption{Feature-level normalized SHAP contributions in the within-subject multi-band model. The standard deviation (SD) indicates inter-participant variability. }
\label{tab:shap_featuretype}
\begin{tabular}{lcc}
\hline
Feature type & Mean norm.\ $|\mathrm{SHAP}|$ & SD \\
\hline
\textbf{Inter-ROI connectivity ($25\%$)}        & \textbf{0.282} & \textbf{0.187} \\
\quad Connectivity                  & 0.271 & 0.188 \\
\quad Hemispheric asymmetry            & 0.006 & 0.009 \\
\quad Anterior-posterior gradient      & 0.004 & 0.006 \\[3pt]

\textbf{Intra-ROI temporal-dynamics ($30\%$)}   & \textbf{0.199} & \textbf{0.128} \\
\quad ROI window low-order statistics                    & 0.153 & 0.092 \\
\quad ROI temporal dynamics            & 0.045 & 0.069 \\[3pt]

\textbf{Intra-ROI spatial-variability ($30\%$)} & \textbf{0.499} & \textbf{0.180} \\
\quad ROI spatial sd.                  & 0.139 & 0.064 \\
\quad ROI spatial range                & 0.135 & 0.059 \\
\quad ROI low-order statistics         & 0.222 & 0.166 \\
\quad ROI high-order statistics        & 0.004 & 0.007 \\[3pt]

\textbf{Global low-order statistics ($15\%$)}                        & \textbf{0.020} & \textbf{0.013} \\
\hline
\end{tabular}
\end{table}

\begin{table}[!tb]
\renewcommand{\arraystretch}{1.3}
\centering
\caption{ROI-level normalized SHAP contributions in the within-subject multi-band model. The standard deviation (SD) indicates inter-participant variability. All regions contain the same number of features.}
\label{tab:shap_region}
\begin{tabular}{lcc}
\hline
Region            & Mean norm.\ $|\mathrm{SHAP}|$ & SD \\
\hline
\textbf{Right prefrontal}        & 0.132 & 0.147 \\
\textbf{Left prefrontal   }      & 0.127 & 0.090 \\
\textbf{Midline frontal  }       & 0.118 & 0.126 \\
\textbf{Left frontal   }         & 0.114 & 0.127 \\
\textbf{Right frontal }          & 0.091 & 0.085 \\
Right centro-temporal   & 0.060 & 0.094 \\
Left parietal           & 0.044 & 0.094 \\
Right fronto-central    & 0.042 & 0.033 \\
Right parietal          & 0.042 & 0.034 \\
Right occipital         & 0.037 & 0.056 \\
Left fronto-central     & 0.037 & 0.028 \\
Left parietal2          & 0.035 & 0.043 \\
Left occipital          & 0.034 & 0.057 \\
Right parietal2         & 0.033 & 0.053 \\
Left centro-temporal    & 0.028 & 0.030 \\
Midline parietal        & 0.026 & 0.021 \\
\hline
\end{tabular}
\end{table}

Table \ref{tab:shap_featuretype} summarizes feature-type–level importance. Intra-ROI spatial-variability features show the largest contributions, indicating that spatial distribution patterns within regions are strongly utilized by the model. Inter-ROI connectivity features represent the second most important group, suggesting that relationships between regions also contribute substantially. Intra-ROI temporal-dynamics features provide complementary information, while global low-order statistics play a relatively minor role.

At the ROI level (Table \ref{tab:shap_region}), frontal and prefrontal regions show the highest contributions, whereas parietal and occipital regions exhibit smaller values. These patterns suggest that the model relies more strongly on distributed frontal activity. However, as with the band-level analysis, these observations reflect model behavior within the given feature space and should not be interpreted as direct evidence of causal neural mechanisms.

\begin{figure}[!tb]
\centerline{\includegraphics[width=3cm]{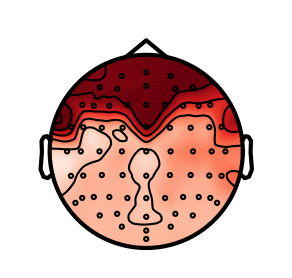}}
\caption{SHAP-based topographic patterns for the multi-band setting}
\label{fig_shaptopo}
\end{figure}

\section{Discussion}
This study examines task-constrained self-initiated (TCSI) shifts using a data-driven framework based on preparatory EEG activity. By focusing on a fixed pre-shift interval, the results show that within-subject classification can distinguish self-initiated from externally instructed shifts under identical visual stimulation. This suggests that the pre-shift period contains structured information that supports subject-specific discrimination within the controlled experimental setting of the adopted paradigm.

The band-specific topographic analysis and SHAP results provide a consistent picture of how different frequency components are utilized by the model. Higher-frequency bands, particularly gamma and high beta, show larger contributions under the current feature representation, and the ROI-level results further emphasize frontal and prefrontal regions. This high-frequency frontal pattern is also compatible with residual frontal EMG activity or preparatory oculomotor signals, especially because shift onset was defined using eye-tracking. Therefore, the present findings should not be interpreted as evidence that high-frequency neural oscillations are the primary mechanism underlying TCSI–EI discrimination. Rather, they show that the EEG-recorded pre-shift signal contains subject-specific discriminative information, which may include both neural and non-neural components. In addition, the feature design and model structure may favor certain frequency components, so the results primarily reflect discriminative utility within the present framework.

The difference from prior studies that emphasize frontal-midline theta activity \cite{HouTheta,ClementsThetaAlpha} can be explained by methodological factors. Previous work typically focuses on specific bands under hypothesis-driven settings, whereas the present approach considers a high-dimensional feature space that combines multiple bands, spatial regions, and connectivity measures within a restricted time window. In addition, the experimental paradigm used here explicitly separates internally and externally driven shifts under identical stimulus conditions, which may lead to different patterns compared to traditional cue-based designs. The present findings are therefore better viewed as complementary to existing results rather than conflicting with them.

The SHAP-based analysis further shows how different feature types and regions are involved in the model. Intra-ROI spatial-variability features contribute the most, followed by inter-ROI connectivity and intra-ROI temporal-dynamics, indicating that both spatial patterns and inter-regional relationships are important. At the regional level, frontal and prefrontal areas show higher contributions than posterior regions. These patterns reflect how the model uses distributed information across regions, and should be interpreted as properties of the learned representation.

Several limitations should be noted. The analysis is based on a single dataset derived from this experimental paradigm, with a limited number of participants, which may limit generalizability. Cross-subject decoding remained at chance level, indicating substantial inter-individual variability and suggesting that subject-specific modeling is important for applications such as personalized and asynchronous brain–machine interface systems. The framework also relies on hand-crafted features and a traditional classifier, which supports interpretability but may limit performance compared to more advanced approaches. Finally, although feature selection was nested within cross-validation, the limited trial counts, high-dimensional handcrafted features, and three-fold cross-validation may still increase the risk of overfitting. In addition, residual non-neural signals, particularly high-frequency EMG or oculomotor components, cannot be completely excluded despite preprocessing \cite{HouPreprocessing}.

\section{Conclusion}
This study presents a subject-specific framework for analyzing preparatory EEG activity associated with task-constrained self-initiated and externally instructed attention shifts. Building on a structured experimental paradigm that enables controlled comparison between internally and externally driven attention, the approach combines topographic analysis with SHAP-based feature attribution to examine how spectral, spatial, and feature-level components are used in within-subject classification. The results show that preparatory EEG activity contains sufficient information for reliable subject-specific discrimination within this paradigm, and that both spatial patterns and inter-regional relationships contribute to model performance. While higher-frequency components show strong contributions, these findings should be interpreted as properties of the feature representation rather than direct indicators of underlying signal patterns.

Overall, this work demonstrates how interpretable machine learning can be used to analyze subject-specific EEG signal patterns in a structured experimental setting. The proposed framework may support the development of personalized and asynchronous BMI systems. Future work will focus on improving cross-subject generalization and include more explicit control analyses to disentangle neural and non-neural contributions in high-frequency bands.

\addtolength{\textheight}{-12cm}   % This command serves to balance the column lengths

%%%%%%%%%%%%%%%%%%%%%%%%%%%%%%%%%%%%%%%%%%%%%%%%%%%%%%%%%%%%%%%%%%%%%%%%%%%%%%%%

%%%%%%%%%%%%%%%%%%%%%%%%%%%%%%%%%%%%%%%%%%%%%%%%%%%%%%%%%%%%%%%%%%%%%%%%%%%%%%%%

%%%%%%%%%%%%%%%%%%%%%%%%%%%%%%%%%%%%%%%%%%%%%%%%%%%%%%%%%%%%%%%%%%%%%%%%%%%%%%%%

%%%%%%%%%%%%%%%%%%%%%%%%%%%%%%%%%%%%%%%%%%%%%%%%%%%%%%%%%%%%%%%%%%%%%%%%%%%%%%%%

\end{document}